\documentclass[10pt,conference]{IEEEtran}

\newcommand{\toolname}{\textrm{\textsc{Gamma}}}

\usepackage[normalem]{ulem}

\newcommand{\revise}[1]{{\color{black}{#1}}}
\newcommand{\delete}[1]{}

\usepackage{tablefootnote}

\usepackage{enumerate}
\usepackage{ragged2e}

\usepackage{stfloats}
\usepackage{cite}
\usepackage{amsmath,amssymb,amsfonts}
\usepackage{algorithmic}
\usepackage{graphicx}
\usepackage{textcomp}

\usepackage{url}
\usepackage{hyperref}
\hypersetup{
    colorlinks=true,
    linkcolor=blue,
    filecolor=blue,      
    urlcolor=blue,
    citecolor=blue,
}

\usepackage{multirow}
\usepackage[utf8]{inputenc}
\usepackage{amsmath,amssymb,amsfonts}
\usepackage{algorithmic}
\usepackage{algorithm}
\usepackage{graphicx}
\usepackage{textcomp}
\usepackage{color}
\usepackage{xcolor}

\usepackage{graphicx}
\usepackage{subfigure}
\usepackage{stmaryrd}
\usepackage{verbatimbox}
\usepackage{enumerate}
\usepackage[shortlabels]{enumitem}
\usepackage{bm}%

\usepackage{caption}

\usepackage{makecell}
\usepackage{booktabs}

\usepackage{multirow}

\usepackage{enumitem}
\usepackage{colortbl}

\usepackage{listings}
  
\lstdefinelanguage{pattern}{
  language=Java,
    backgroundcolor=\color{gray!10!white},   
    commentstyle=\color{green},
    keywordstyle=\color{red},
    numberstyle=\tiny\color{gray},
    stringstyle=\color{red},
    basicstyle=\ttfamily\footnotesize,
    breakatwhitespace=false,         
    breaklines=true,                 
    captionpos=b,                    
    keepspaces=true,                 
    numbersep=5pt,                  
    showspaces=false,                
    showstringspaces=false,
    showtabs=false,                  
    tabsize=2,
    morecomment=[s][\color{green!80!blue}]{+\ },
    morecomment=[s][\color{red}]{-\ },
    morecomment=[n][\color{violet}]{<mask}{>},
    escapeinside={(*}{*)},
}

\lstdefinelanguage{diff}{
    language=Java,
    backgroundcolor=\color{white}, 
    commentstyle=\color{green},
    keywordstyle=\color{black},
    numberstyle=\tiny\color{gray},
    stringstyle=\color{red},
    basicstyle=\ttfamily\footnotesize,
    breakatwhitespace=false,         
    breaklines=true,                 
    captionpos=b,                    
    keepspaces=true,                 
    numbers=left,                    
    numbersep=5pt,                  
    showspaces=false,                
    showstringspaces=false,
    showtabs=false,                  
    tabsize=2,
    morecomment=[l][\color{gray}]{@@},
    morecomment=[l][\color{green!70!black}]{+\ },
    morecomment=[l][\color{red}]{-\ },
    frame = TB,
}

\usepackage{xspace}

\AtBeginDocument{%
  \providecommand\BibTeX{{%
    \normalfont B\kern-0.5em{\scshape i\kern-0.25em b}\kern-0.8em\TeX}}}
    
\begin{document}
\IEEEoverridecommandlockouts %

\title{{\toolname}: Revisitin\underline{g} Template-based \underline{A}utomated Progra\underline{m} Repair via \underline{Ma}sk Prediction}

\author{Anonymous Author(s)}

\author{
\IEEEauthorblockN{Quanjun Zhang}
\IEEEauthorblockA{
State Key Laboratory for Novel \\ Software Technology \\
Nanjing University, China\\
quanjun.zhang@smail.nju.edu.cn}

\\
\IEEEauthorblockN{Bowen Yu}
\IEEEauthorblockA{
State Key Laboratory for Novel \\ Software Technology \\
Nanjing University, China\\
201250070@smail.nju.edu.cn}

\and
\IEEEauthorblockN{Chunrong Fang\IEEEauthorrefmark{1} \thanks{\IEEEauthorrefmark{1}Chunrong Fang and Zhenyu Chen are the corresponding authors.}}
\IEEEauthorblockA{
State Key Laboratory for Novel \\ Software Technology \\
Nanjing University, China\\
fangchunrong@nju.edu.cn}

\\
\IEEEauthorblockN{Weisong Sun}
\IEEEauthorblockA{
State Key Laboratory for Novel \\ Software Technology \\
Nanjing University, China\\
weisongsun@smail.nju.edu.cn}

\and
\IEEEauthorblockN{Tongke Zhang}
\IEEEauthorblockA{
State Key Laboratory for Novel \\ Software Technology \\
Nanjing University, China\\
201250032@smail.nju.edu.cn}

\\
\IEEEauthorblockN{Zhenyu Chen\IEEEauthorrefmark{1}}
\IEEEauthorblockA{
State Key Laboratory for Novel \\ Software Technology \\
Nanjing University, China\\
zychen@nju.edu.cn}
}

\maketitle

\begin{abstract}
Automated program repair (APR) aims to fix software bugs without human intervention and plays a crucial role in software development and maintenance.
Template-based APR has been widely investigated and shown promising results.
However, it is challenging for template-based APR to select the appropriate donor code, which is an important repair ingredient for generating candidate patches.
Inappropriate donor code may cause plausible but incorrect patch generation even with correct fix patterns, limiting the repair performance.

In this paper, we aim to revisit template-based APR, and propose {\toolname}, to directly leverage large pre-trained language models for donor code generation.
Our main insight is that instead of retrieving donor code in the local buggy file, we can directly predict the correct code tokens based on the
context code snippets and repair patterns by a cloze task.
Specifically,
(1) {\toolname} revises a variety of fix templates from state-of-the-art template-based APR techniques (i.e., TBar) and transforms them into mask patterns.
(2) {\toolname} adopts a pre-trained language model to predict the correct code for masked code as a fill-in-the-blank task.
Although our idea is general and can be built on various existing pre-trained language models, we have implemented {\toolname} as a practical APR tool based on the recent UniXcoder model.
The experimental results demonstrate that {\toolname} correctly repairs 82 bugs on Defects4J-v1.2, which achieves 20.59\% (14 bugs) and 26.15\% (17 bugs) improvement over the previous state-of-the-art template-based approach TBar and learning-based one Recoder.
Furthermore, {\toolname} repairs 45 bugs and 22 bugs from the additional Defects4J-v2.0 and QuixBugs, indicating the generalizability of {\toolname} in addressing the dataset overfitting issue.
We also prove that adopting other pre-trained language models can provide substantial advancement, e.g., CodeBERT-based and ChatGPT-based {\toolname} is able to fix 80 and 67 bugs on Defects4J-v1.2, indicating the scalability of {\toolname}.
Overall, our study highlights the promising future of adopting pre-trained models to generate correct patches on top of fix patterns in practice.
  
\end{abstract}

\begin{IEEEkeywords}
Automated Program Repair, Fix Pattern, Pre-trained Model, LLM4SE
\end{IEEEkeywords}

\section{Introduction}
\label{sec:intro}
The complexity and size of modern software systems are continuously enlarging, leading to the soaring number of software bugs~\cite{monperrus2018automatic,hao2015optimal}. 
Software bugs have detrimental effects on software development, as they give users an annoying experience, and sometimes can cause huge financial losses to developers~\cite{gazzola2019automatic}. 
A considerable amount of time and budget is spent on identifying and fixing such software bugs manually \cite{benton2020effectiveness}.
To facilitate the process of manual debugging, automated program repair (APR), which aims to generate correct patches for identified buggy code snippets automatically, is getting growing attention from both academia and industry~\cite{gao2022program,zhang2023survey}, such as Meta \cite{marginean2019sapfix}, Google \cite{mesbah2019deepdelta} and Microsoft \cite{drain2021generating, allamanis2021self}. 

In the literature, a variety of APR techniques have been proposed to generate patches, such as heuristic-based~\cite{ Martinez2016Astor, Yuan2018Arja}, constraint-based~\cite{Durieux2016Dynamoth, Xuan2016Nopol}, template-based \cite{koyuncu2020fixminer, Liu2019Avatar}.
Among these traditional APR techniques, template-based APR, which employs repair patterns hand-crafted by human experts to transform buggy code snippets into correct ones,  has been widely investigated and recognized as state-of-the-art \cite{liu2019tbar,xia2022less,xia2023automated}.
Candidate patches are usually generated by leveraging two kinds of repair ingredients (i.e., fix patterns and donor code) that are found in existing code bases.
The repair pattern represents common code change actions (e.g., insertion of an If statement) 
and donor code represents code fragments (e.g., identifier tokens such as method names) to concretize patches guided by abstract patterns.
A mass of studies has been dedicated to template extraction schemes, such as  manually extracted templates and automatically mining templates \cite{liu2018mining,wen2018context,jiang2018shaping,hua2018towards}.
For example, state-of-the-art template-based APR tool TBar~\cite{liu2019tbar} focuses on the local buggy file and leverages the context of buggy code to prune away irrelevant donor code.
Previous works~\cite{yang2021were,liu2023reliable} have shown a considerable number of bugs cannot be fixed because the relevant donor code is not available in the local file.
Therefore, \delete{Tbar}\revise{TBar} may fail to generate correct patches with inappropriate donor code although the fix pattern matches with correct code change actions.

In this paper, we propose a novel template-based APR tool called {\toolname} by
combining the advances of fix patterns and pre-trained language models.
The key insight is that considering pre-trained models can acquire adequate general knowledge about programming language from all possible open-source projects in the wild, we can directly employ such models to retrieve relevant donor code from the fix pattern and surrounding code context.
In particular, we first collect and summarize a super-set of fix patterns drawn from previous template-based work (e.g., \delete{Tbar}\revise{TBar}).
We then transform these fix templates into hole-filling-based patterns, which replace donor code with several masked tokens to be filled.
Finally, we perform a mask prediction task on the hole-filling-based fix patterns with the help of pre-trained models in a fill-in-the-blank manner,
i.e., predicting the correct donor code for the masked tokens.
Although {\toolname} is conceptually generalizable to various pre-trained models, we have implemented {\toolname} on top of one recent pre-trained language model, UniXcoder~\cite{guo2022unixcoder}.
Unlike current template-based APR tools which usually retrieve fix ingredients in the local buggy file, {\toolname} directly utilizes generic knowledge pre-trained with millions of code snippets from open-source projects, allowing it to provide a variety of donor code to fix different bugs.

We conduct extensive experiments to compare {\toolname} with state-of-the-art APR approaches (including both traditional and learning-based ones) on the widely-adopted Defects4J-v1.2 benchmark.
The experimental results demonstrate that {\toolname} is able to outperform all existing APR approaches, improving the number of correctly-fixed bugs to 82 with a precision of 81.19\%, and 14 unique bugs that no prior work can fix, which is a new frontier in the APR field.
Besides, {\toolname} fixes 45 and 22 bugs on the additional Defects4J-v2.0 and QuixBugs, 27 and 5 more than state-of-the-art learning-based technique Recoder, demonstrating that {\toolname} can address the important dataset overfitting issue well.
Moreover, we implement {\toolname} with CodeBERT~\cite{feng2020codebert} and ChatGPT~\cite{chatgpt}, and find 80 and 67 bugs are fixed correctly on Defects4J-v1.2.
The results demonstrate that {\toolname} with other pre-trained models can further provide substantial advancement, highlighting the generalizability of {\toolname}.

To sum up, the contributions of this paper are as follows:
\begin{itemize}
  \item \textbf{New Dimension.}
  We bridge the gap between the advances in recent pre-trained models and template-based APR.
  Different from existing template-based APR retrieving donor code from local buggy files and existing learning-based APR generating a patched code snippet from scratch, our work demonstrates that we can leverage pre-trained models to generate correct code tokens in a given fix pattern. 
  More importantly, our work reveals the potential for leveraging pre-trained models to resolve the important fix ingredient problem in template-based APR.

  \item \textbf{Novel APR tool.}
  We propose {\toolname}, which leverages the large pre-trained language model to generate correct code with the help of fix patterns without any additional historical bug-fixing pairs for training.
  We define a set of fix patterns in a fill-in-the-blank format and leverage the original pre-training objective of pre-trained models to predict actual masked-out tokens.
  Considering the fill-in-the-blank task can leverage various pre-trained language models, {\toolname} is general in concept and can be implemented with different pre-trained models in practice.

  \item \textbf{Extensive study.}
  We conduct an empirical study to investigate the effectiveness of {\toolname} compared to state-of-the-art traditional and learning-based APR techniques.
  The results on the widely-adopted Defects4J-v1.2 show that {\toolname} is able to fix 82 bugs and 14 of them cannot be fixed by existing APR tools, creating a new higher baseline of repair performance.
  More importantly, {\toolname} fixes 45 and 22 bugs on the newly-developed Defects4J-v2.0 and QuixBugs, demonstrating that {\toolname} can avoid the important dataset overfitting issue of existing APR techniques.
  Moreover, we adopt different pre-trained models (e.g., ChatGPT) to further investigate the generalization ability of {\toolname}.

  \item \textbf{Available artifacts.}
  To support the open science community, we release the relevant materials (including source code, experimental results, and correct patches) in our experiment for replication and future research \cite{myurl}.
  
\end{itemize}

\section{Background and Motivation}
\subsection{Automated Program Repair}
As a promising technique to shift the
heavy manual debugging to efficient automated patch generation, APR has developed rapidly and received much attention from a broad of research communities, such as software engineering, software security, and artificial intelligence \cite{zhang2023survey,yang2022survey}.
The workflow of APR usually involves three phases:
(1) \textit{fault localization}, i.e., the off-the-shelf fault localization techniques are utilized to identify a ranked list of suspicious code elements, with whose help APR can focus on a small code region, thus reducing the workload~\cite{wong2016survey};
(2) \textit{patch generation}, i.e., candidate patches are generated by applying a set of transformation rules to the suspicious code snippets~\cite{liu2020efficiency};
and (3) \textit{patch validation}, i.e., the available test suites are utilized as the program specifications to check the correctness of candidate patches~\cite{zhang2023boosting}.
The candidate patches that pass all available test suites are considered plausible ones.
The plausible patches that are semantically equivalent to the
developer patch by manual inspection are considered correct ones; otherwise overfitting ones ~\cite{benton2022towards,zhang2022interactive}.

In the literature, as the core component of APR research, a mass of research efforts are devoted to generating patches from different aspects, including traditional and learning-based ones.
In particular, traditional APR techniques can be classified as heuristic-based~\cite{ Martinez2016Astor, Yuan2018Arja}, constraint-based~\cite{Durieux2016Dynamoth, Xuan2016Nopol}, template-based \cite{koyuncu2020fixminer, Liu2019Avatar}.
Among them, template-based APR is proven to achieve the best performance, which consists of two fix ingredients, i.e., fix patterns and donor code.
Fix patterns are hand-crafted by human experts to denote the common code changes, and donor code is retrieved in buggy files to denote the actual correct code tokens.
{\toolname} aims to revise the important donor code by employing pre-trained language models in a fill-in-the-blank manner.

Compared with traditional APR techniques, learning-based techniques handle the program repair problem as a neural machine translation (NMT) task, which translates a code sequence from a source language (i.e., buggy code snippets) into a target language (i.e., correct code snippets).
Existing NMT repair models are typically built on the top of the \textit{encoder-decoder} architecture \cite{vaswani2017attention}.
The \textit{encoder} extracts the hidden status of buggy code snippets with the necessary context, and the \textit{decoder} takes the encoder's hidden status and generates the correct code snippets~\cite{long2016automatic,jiang2018shaping, li2020dlfix}.
Thanks to the powerful ability of DL to learn hidden and intricate relationships from massive code corpora, learning-based APR techniques have achieved remarkable performance in the last couple of years.
Although learning-based APR techniques have demonstrated their promising future, they are still limited by the quality and quantity of historical bug-fixing pairs for training~\cite{xia2022less}.
We view {\toolname} as a novel learning-based APR technique that attempts to boost traditional APR techniques by utilizing deep learning technology.
However, different from most existing learning-based APR that treats patch generation as an end-to-end NMT task with a limited number of bug-fixing pairs as training data, {\toolname} integrates pre-trained language models into template-based APR and only predicts masked code tokens with a zero-shot learning scenario.

\subsection{Pre-trained Model}

Recently, pre-trained language models (e.g., UniXcoder~\cite{guo2022unixcoder} and ChatGPT~\cite{chatgpt}) have significantly boosted performance across a wide range of code-related tasks~\cite{yuan2022circle,li2022automating}.
These models are pre-trained by self-supervised training on large-scale unlabeled corpora and then fine-tuned by supervised training on limited corpora to enhance performance on multiple downstream tasks.
During the pre-training process, a masked language modeling objective is usually employed to derive generic language representations from the massive unlabeled training data~\cite{devlin2018bert}, i.e., a small percentage of tokens are replaced by masked tokens, and the training objective is to predict the original values of the masked tokens.

Existing pre-trained models usually adopt the encoder-decoder architectures, where the former encodes an input sequence as a fixed-length vector representation, and the latter generates an output sequence based on the input representation.
These models can be generally categorized into three architectures: encoder-only, decoder-only, and encoder-decoder ones \cite{niu2022spt}. 
Encoder-only models (e.g., CodeBERT \cite{devlin2018bert}) usually pre-train a bidirectional transformer in which each token can attend to each other.
Decoder-only models (e.g., GPT \cite{brown2020gpt}) are pre-trained using unidirectional language modeling that only allows tokens to attend to the previous tokens and themselves to predict the next token.
Encoder-decoder models (e.g., UniXcoder~\cite{guo2022unixcoder}) often make use of denoising pre-training objectives that corrupt the source input and require the decoder to recover them.

In this work, we select UniXcoder to retrieve the doner code via a mask prediction task.
UnixCoder is pre-trained using the MLM objective which can be used to directly generate masked code tokens from the appropriate fix pattern and surrounding code context.
Besides,  CodeBERT and ChatGPT are employed to investigate the generalization ability of {\toolname}.

\subsection{Fix Template}
Fix templates are widely employed in the APR community~\cite{liu2019tbar}. 
A fix template is a pre-defined code transformation rule that represents a common code change in the bug-fixing process. 
The insight behind fix templates is that many software bugs are similar in nature~\cite{jiang2018shaping}. 
Therefore, with fix templates summarized from previous bugs, it is possible to automatically fix some other flawed code~\cite{yang2021were}.

In the literature, there are several strategies to access fix templates: 
(1) \textit{manual template mining}~\cite{kim2013automatic,pan2009toward}, i.e.,  through performing analysis on existing bugs as well as their relevant patches, experienced developers can identify similar code changes, and turn them into fix templates. 
(2) \textit{machine learning}~\cite{liu2018mining,koyuncu2020fixminer}, i.e., learning approaches are used so that fix templates can be generated automatically.
(3) \textit{static analysis}~\cite{Liu2019Avatar,liu2018mining2}, i.e.,  fix templates are generated from assorted types of warnings raised by static analysis tools.

The process of applying a fix template to patch generation typically involves two steps. 
First, the APR tool selects an appropriate fix template based on the abstract syntax tree (AST) representation of the buggy code.
Fix templates are chosen according to the types of nodes in AST, and bugs are fixed by mutating the target nodes.
Second, the APR tool generates a repaired version of the buggy code by searching and applying the relevant donor code to the fix pattern.

\begin{lstlisting}[float,floatplacement=H,language=diff,caption={The Defects4J bug Closure-92},label={listing:closure92}]
@@ -786,7 +786,7 @@ void replace() {
    } else {
-   int indexOfDot = namespace.indexOf('.');
+   int indexOfDot = namespace.lastIndexOf('.');
    if (indexOfDot == -1) {            
        compiler.getNodeForCodeInsertion(minimumModule)
\end{lstlisting}

\subsection{Motivation Example}
\label{sec:moti}
To better illustrate the limitation of existing template-based APR, we further present a motivating example in this section.
As shown in Listing~\ref{listing:closure92}, we use a real-world bug Closure-92 from the widely-used benchmark Defects4J-v1.2 as an example.
Closure-92 denotes the 92nd buggy version of the Google Closure Compiler project in Defects4J-v1.2.
This bug is fixed by {\toolname} successfully, while \delete{Tbar}\revise{TBar} fails to generate a correct patch.
To fix this bug, the method name ``indexOf'' is replaced by ``lastIndexOf''. 
The fix template used here is to mutate method names. 
TBar applies the selected fix patterns to the source code in a naive way. 
In this case, TBar searches all the methods that appear in the local file where the bug is localized, and replaces the buggy method with all the other methods with the same return type one by one. 
As a result, TBar is not able to generate method names that do not exist in the original file, like the ``lastIndexOf'' in this example, limiting its repair performance.
Different from TBar, we replace the method name ``indexOf'' with a mask token (i.e. $<$mask$>$) instead, and query the pre-trained model UniXcoder to fill the mask with the fix pattern and corresponding context.

Based on the example, we can observe that, although the correct fix pattern is selected, as a state-of-the-art traditional APR, \delete{Tbar}\revise{TBar} still fails to generate a correct patch with inappropriate donor code (i.e., ``lastIndexOf'').
The effectiveness of template-based APR is largely dependent on donor code, which refers to the code tokens (e.g., variable name) that can be combined with the fix template in order to produce a complete patch. 
Donor code can be accessed in different scopes of the buggy program (e.g. a method, a file, or a package), but for some bugs, the correct donor code cannot be found even if the whole program is searched.
For example, previous work demonstrates that half of the bugs from the Defects4J benchmark cannot be fixed because the relevant donor code is not available in the search space~\cite{yang2021were}.
Thus, these bugs cannot be correctly fixed by template-based APR tools like TBar\cite{liu2019tbar}, which only identifies the donor code within the local file. 
With a larger search space (e.g., searching donor code from other projects), there might be more chances to fix these bugs.
However, such a strategy leads to the search space explosion problem and an unaffordable search time, reducing repair efficiency.
In this paper, we utilize pre-trained language models to retrieve relevant donor code.
These models have learned programming language knowledge from a great number of programs in the wild, making it possible to repair bugs that require donor code from outside the buggy program.

\section{Approach}
\begin{figure*}[ht]
    \centering
    \includegraphics[width=0.98\linewidth]{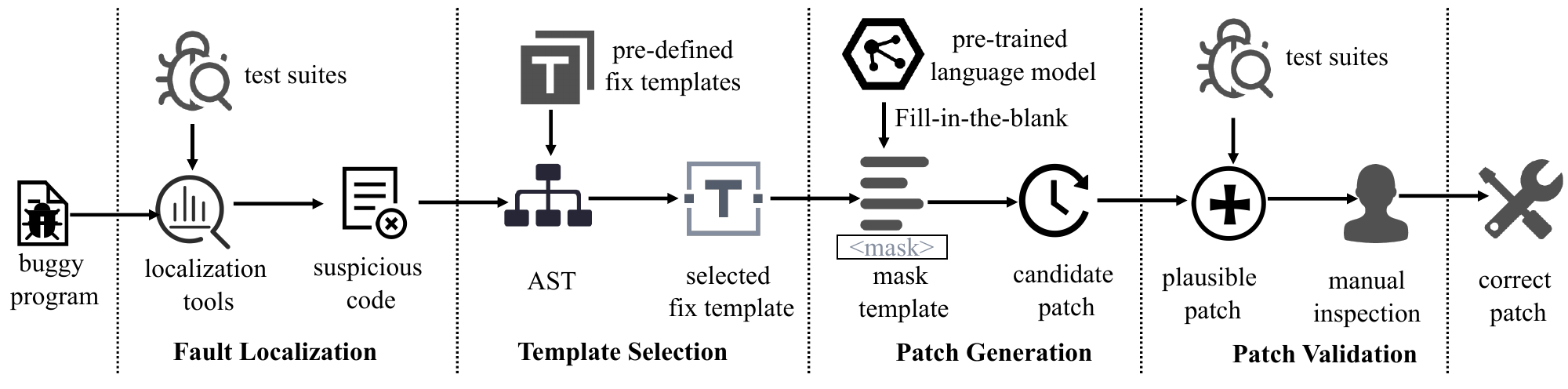}
    \caption{The overall workflow of {\toolname}}
    \label{fig:workflow}
\end{figure*}

In order to assess the effectiveness of fix ingredients,  we build {\toolname}, a template-based APR tool that combines the recurrently-used fix patterns and pre-trained language models.
Fig.~\ref{fig:workflow} presents the workflow of {\toolname}.
Given a buggy program and a set of test suites that make the program fail, a list of suspicious code elements is returned by fault localization approaches (i.e., \textit{fault localization phase}).
On top of the existing fix pattern corpus, {\toolname} then selects appropriate fix patterns to the suspicious elements (i.e., \textit{pattern selection phase}) and queries pre-trained models to retrieve donor code via a mask prediction mask (i.e., \textit{patch generation phase}).
{\toolname} finally employs the available test suites as the oracle to check the generated patches and returns plausible patches for manual inspection (i.e., \textit{patch validation phase}).
Considering that fault localization is usually developed as an independent field and existing APR techniques employ off-the-shelf fault localization tools in the repair pipeline, we do not discuss the fault localization below.
We describe the role and operation of other phases as well as all necessary implementation details.

\subsection{Mask Template Definition}
In the literature, a variety of fix patterns are designed based on manual summarization or automatic mining.
On top of the state-of-the-art template-based approach TBar~\cite{liu2019tbar}, we manually inspect all fix patterns and transform them into hole-filling-based patterns.
We show the related templates as well as how they are applied to buggy code, or how statements with mask tokens are generated based on these templates.

{\textit{T1: Check Cast Expressions.}}
Adding an instanceof check around a statement when it contains an unchecked cast expression.
\begin{lstlisting}[language=pattern]
+   if(exp instanceof T){
        var=(T) exp ...
+   }
\end{lstlisting}

\textit{T2: Mutate Conditional Expressions:}
Mutating an expression that returns a boolean value by removing part of the expression, replacing it with masks or adding new masks.
\begin{lstlisting}[language=pattern]
(*\bfseries Removing expression: *)
-   condExp1 op condExp2
+   condExp1

(*\bfseries Updating expression: *)
-   condExp1 op condexp2
+   condExp1 op <mask>

(*\bfseries Adding expression: *)
-   condExp1
+   condExp1 <mask>
\end{lstlisting}
where {\em condExp} denotes conditional expressions and {\em Op} denotes the logical operator (i.e., $\|$ or $\&\&$).

\textit{T3: Mutate Data Types:}
Using one or more masks to replace data types in the variable declaration or cast expression nodes.
\begin{lstlisting}[language=pattern]
-   T var = ...
+   <mask> var = ...

-   ... (T) exp ...
+   ... (<mask>) exp ...
\end{lstlisting}
where both {\em T} denotes a data type and {\em exp} denotes the being casted expression (e.g., variable).

\textit{T4: Mutate Literal Expressions:}
Replacing literal expressions, including number literals, string literals, boolean literals, etc. with masks.
\begin{lstlisting}[language=pattern]
-   ...literal...
+   ...<mask>...
\end{lstlisting}

\textit{T5: Mutate Method Invocations.}
Mutating method invocation expressions by changing either method names or arguments.

\begin{lstlisting}[language=pattern]
(*\bfseries Method name replacement: *)
-   method(...)
+   <mask>(...)

(*\bfseries Argument insertion: *)
-   method(arg)
+   method(arg,<mask>)

(*\bfseries Argument removal: *)
-   method(arg1,arg2)
+   method(arg)

(*\bfseries Argument replacement: *)
-   method(arg)
+   method(<mask>)
\end{lstlisting}

{\textit{T6: Check Null Pointer:}}
Adding a null check to a statement that contains an expression that is probably null.
\begin{lstlisting}[language=pattern]
(*\bfseries Null point skip: *)
+   if (exp != null){
        ...exp...
+   }

(*\bfseries Return insertion: *)
+   if (exp == null){
+       return <mask>;
+   }
    ...exp...
    
(*\bfseries Continue: *)
+   if (exp == null){
+       continue;
+   }
    ...exp...
    
(*\bfseries Exception throw: *)
+   if (exp == null){
+       throw new IllegalArgumentException();
+   }
    ...exp...

(*\bfseries Re-assignment: *)
+   if (exp == null){
+       exp=<mask>
+   }
    ...exp...
\end{lstlisting}

\textit{T7: Mutate Operators:}
Replacing an operator in a statement with masks or changing the priority of operations.
\begin{lstlisting}[language=pattern]
(*\bfseries Changing the priority: *)
-   (exp1 op1 exp2) op2 exp3
+   exp1 op1 (exp2 op2 exp3)

(*\bfseries Replacing operator: *)
-   exp1 op exp2
+   exp1 <mask> exp2
\end{lstlisting}

\textit{T8: Check Array Range:}
Checking the range of index before accessing an element in an array.
\begin{lstlisting}[language=pattern]
+   if (index<array.length) {
        ...array[index]...
+   }
\end{lstlisting}

\textit{T9: Mutate Return Statements:}
Replacing the expression (e.g., literals, variables, and conditional expressions) that is returned in a method with masks.
\begin{lstlisting}[language=pattern]
-   return exp;
+   return <mask>;
\end{lstlisting}

\textit{T10: Insert Statements:}
Inserting return statements, try catch statements, if statements, method invocations, or simply some masks to the existing statement.
\begin{lstlisting}[language=pattern]
(*\bfseries Return statement: *)
+   return <mask>;
    statement;
    
(*\bfseries Try-catch statement: *)
+   try{
        statement;
+   } catch(Exception e){}

(*\bfseries If statement: *)
+   if (<mask>) {
        statement;
+   }

(*\bfseries Simple statement: *)
+   <mask>;
    statement;
    
\end{lstlisting}

\textit{T11: Remove Statements:}
Directly deleting one or more buggy statements from the original code.
\begin{lstlisting}[language=pattern]
-   statement;
\end{lstlisting}

\textit{T12: Replace Variables:}
Replacing a variable in a buggy statement with masks.
\begin{lstlisting}[language=pattern]
-   ...var...
+   ...<mask>...
\end{lstlisting}

\textit{T13: Move statements:}
Moving a statement from its original position to a new one.
\begin{lstlisting}[language=pattern]
-   statement;
    ...
+   statement;
\end{lstlisting}

\subsection{Template Selection}

On top of the defined mask fix templates, {\toolname} determines which template should be applied to the input buggy statements.
Similar to TBar~\cite{liu2019tbar}, {\toolname} employs an AST-based matching approach with a depth-first strategy.
We first generate an AST for the input, and then all the nodes in the AST are traversed. If the AST contains a type of node that is required in a specific template, we apply the template to the bug.
We give an example of the \emph{Closure-10} bug from the Defects4J-v1.2 in Listing \ref{listing:closure10} and how we generate its patch in Fig.~\ref{fig:workflow}.
In the AST of the input line, we find a \emph{Method Invocation} node in the AST of the input line, so we choose the template \emph{T5: Mutate Method Invocation}. 
For the template \emph{Method name replacement} that alters the name of a method, after locating the method name \emph{allResultsMatch}, we replace it with a mask token, which is going to be predicted in the next phase.
\revise{It is worth noting that {\toolname} is built on top of UniXCoder, which is able to predict a sequence of code tokens based on a masked token. Thus, we do not need to consider how many tokens should be masked during patch generation and only use one mask token in the selected repair template, which is different from other pre-trained models, such as CodeBERT used in AlphaRepair~\cite{xia2022less} (discussed in Section~\ref{sec:rq3}).
There might be multiple repair templates that are suitable for a piece of buggy code at the same time.
In this case, we stop the selection of repair  templates as soon as the first correct patch is generated.}
\delete{We do not know the exact number of masks that we should use as there could be a great many possibilities in the patch. So while masking, we naively try all the mask numbers from 1 to 20, which is a range suitable to most cases.
The processes of creating masked statements with other templates are similar to what is described above.}

\begin{figure}[htbp]
\begin{lstlisting}[language=diff,caption={The Defects4J bug Closure-10},label={listing:closure10}]
@@ -1414,7 +1414,7 @@ static boolean mayBeString(Node n) {
static boolean mayBeString(Node n, boolean recurse) {
    if (recurse) {
-       return allResultsMatch(n, MAY_BE_STRING_PREDICATE);
+       return anyResultsMatch(n, MAY_BE_STRING_PREDICATE);
    } else {
    return mayBeStringHelper(n);
    }
\end{lstlisting}
\end{figure}

\subsection{Patch Generation with Mask Prediction}
After selecting an appropriate fix template for a buggy code, we use UniXcoder~\cite{guo2022unixcoder} to generate the correct code tokens via a fill-in-the-blank format.
To this end, we leverage the original pre-training training objective of masked language modeling in UniXcoder.
UniXcoder is an advanced pre-trained model for programming languages that support code understanding and generation tasks. 
It contains a pre-training objective of Masked Language Modeling (MLM), which is designed to predict some tokens that have been masked out.
We leverage this pre-training task to complete the clozes generated in the previous step with any fine-tuning, so that candidate patches for the buggy programs can be produced. 

The prediction for the mask is largely dependent on the tokens surrounding the mask. If the correct token appears in the input for the model, it is more likely that the token will be chosen as one of the possible results. 
The precision of mask prediction is quite limited when only a single masked buggy line is given without any context of the code, where there may be some useful information for bug fixing. 
To get more context of the buggy line in each bug, we extract the method that contains the line and use the whole method with the masked buggy line as the input for UniXcoder. Considering that some tokens in the buggy line have been replaced with masks but these tokens may also contain essential information for mask prediction, in the first line of our input, we add the original buggy line in the form of a comment (i.e., add a ``//'' in front of the line). 
The commented buggy line followed by the method the bug is in together forms the final input.
For every input, N candidate patches are generated by the UniXcoder model.
N is the beam size and is an adjustable parameter of {\toolname}.
Relatively large beam size increases the possibility of generating correct patches.

\subsection{Patch Validation}
After a candidate patch for a given bug is generated by {\toolname}, we apply the corresponding changes to the buggy program.
Following the practice in the APR community~\cite{jiang2021cure,xia2022less}, we first recompile the patched program and filter out any patches that fail to compile.
We then execute the patched program against the available test suite to identify the plausible patches that successfully pass all the test suites.
For those plausible patches, we examine them manually to ensure the programs are fixed correctly, i.e., whether the patches are semantically equivalent to developer patches.

\section{Experimental Setup}

\subsection{Research Questions}
In this paper, we study the following research questions:

\textbf{RQ1}: 
What is the performance of {\toolname} compared to state-of-the-art APR approaches?

\textbf{RQ2}: 
What is the generalizability of {\toolname} in repairing additional real-world bugs?

\textbf{RQ3}: 
What is the scalability of {\toolname} when employing other advanced pre-trained models?

\subsection{Benchmarks}
\label{sec:baseline}
To evaluate the repair performance,  we use the standard benchmark of Defects4J-v1.2~\cite{just2014defects4j} in the APR community.
Defects4J-v1.2 is a collection of real-world bugs from open-source projects and is widely adopted by existing traditional~\cite{liu2019tbar,zhang2022program1} and learning-based APR approaches~\cite{yuan2022circle,li2020dlfix,jiang2021cure,zhu2021syntax}. 
In particular, Defects4J-v1.2 contains 395 known and reproducible bugs, each of which contains a buggy version and a fixed version, as well as a corresponding test suite that triggers that bug for patch validation.
Evaluation on Defects4J-v1.2 can reflect the performance of {\toolname} in a real-world debugging scenario and provides sufficient comparison results against most of the existing APR techniques.

Besides, we choose Defects4J-v2.0 and QuixBugs as other bug benchmarks for evaluation, so as to investigate the generalizability of {\toolname}.
\delete{Defects-v2.0}\revise{Defects4J-v2.0} provides 420 additional real-world bugs from 17 Java projects, which is adopted by some recent APR studies~\cite{zhu2021syntax,xia2022less}.
QuixBugs \cite{lin2017quixbugs} is a multi-lingual parallel bug-fixing dataset in Python and Java used in \cite{xia2022less, yuan2022circle}.
QuixBugs contains 40 small classic algorithms with a bug on a single line, along with the bug-triggering test suite.

\subsection{Baselines}
To enable sufficient evaluations, we compare {\toolname} against both traditional and learning-based APR approaches.
We choose seven recent learning-based APR tools, i.e., AlphaRepair~\cite{xia2022less}, Recoder~\cite{zhu2021syntax}, CURE~\cite{jiang2021cure}, CoCoNuT~\cite{lutellier2020coconut}, CIRCLE~\cite{yuan2022circle}, DLFix~\cite{li2020dlfix}, and SequenceR~\cite{chen2019sequencer}.
We also choose two state-of-the-art template-based APR tools TBar~\cite{liu2019tbar} and PraPR~\cite{ghanbari2019practical} as representatives of traditional APR.
In total, we evaluate {\toolname} against nine advanced APR tools from different categories.
Although the fault localization configuration is a significant part of APR, we do not take it into consideration in our experiment because of potential deviations that fault localization may bring about.
Following recent APR studies~\cite{yuan2022circle,li2020dlfix,jiang2021cure,zhu2021syntax}, \revise{we apply perfect fault localization in the way of inputting the exact buggy lines into different APR techniques to standardize the impact of fault localization on repair performance, discussed in Section~\ref{sec:threats}.}

\subsection{Evaluation Metrics}

We use two common metrics to evaluate the performance of all involved APR approaches \cite{gao2022program,zhang2023survey}, i.e., plausible patch and correct patch.
The first one fixes the buggy functionality without harming other correct functionality (i.e., passing all available test suites), and the second one is semantically or syntactically equivalent to the developer patch (i.e., generalizing the potential test suite).
We manually inspect each plausible patch to identify whether it is a correct patch by following the standard practice in APR research.

\subsection{Implementation Details}
At the stage of fix template selection, we apply Eclipse JDT to parse the input line into AST, and then the AST is traversed to examine if it contains any node that is required by a fix template.
There are several templates that can fit all the input buggy lines. 
For example, the template \emph{T10: Insert Statements} only requires adding statements around the buggy line and does not mutate any nodes in the AST. 
Such templates are directly applied to all the inputs without checking the AST.

In the mask prediction phase, we choose the UniXcoder model ``unixcoder-base''. 
This is a model pre-trained on natural language-programming language (NL-PL) pairs and is reported in the original UniXcoder paper \cite{guo2022unixcoder}. 
We use the encoder-decoder mode of the model to give predictions for each mask and generate candidate patches. 
We set the beam size as 250 due to the limitation of our device, which is smaller than 1000 used in CURE~\cite{jiang2021cure} and CoCoNuT~\cite{lutellier2020coconut}.
\revise{Following previous learning-based APR approaches~\cite{zhu2021syntax,xia2022less}, we set a 5-hour running-time limit for fixing one bug to perform a fair comparison.}
w

All experiments are conducted on one Ubuntu 18.04.3 server with two Tesla V100-SXM2 GPUs.

\section{Evaluation and Results}

\subsection{Comparison with State-of-the-arts}

\begin{table*}[htbp]
  \centering
\caption{Comparison with state-of-the-art APR techniques.
\revise{Following the common practice in the APR community~\cite{yuan2022circle,jiang2021cure,lutellier2020coconut}, we reuse the released results from the most recent work \cite{xia2022less} instead of directly running
the APR tools.
Due to the APR community's subsequent validation of publicly released correct patches, the results of some APR tools may be different from the reported results in their published papers.
(\dag) PraPR is evaluated with the results of Ochiai fault localization~\cite{wong2016survey}.
}
}
\label{tab:effectiveness}
    \begin{tabular}{c|cccccccccc}
    \toprule
    Project & SequenceR & CoCoNuT & CURE  & DLFix & Recoder & AlphaRepair & CIRCLE & PraPR\dag & Tbar  & {\toolname} \\
    \midrule
    Chart & 3     & 7     & 10    & 5     & 10    & 9     & 7     & 7     & 11    & \textbf{11} \\
    Closure & 3     & 9     & 14    & 11    & 21    & 23    & 17    & 12    & 16    & \textbf{24} \\
    Lang  & 2     & 7     & 9     & 8     & 11    & 13    & 10    & 6     & 13    & \textbf{16} \\
    Math  & 6     & 16    & 19    & 13    & 18    & 21    & \textbf{27} & 10    & 22    & 25 \\
    Mockito & 0     & 4     & 4     & 1     & 2     & \textbf{5} & 1     & 3     & 3     & 3 \\
    Time  & 0     & 1     & 1     & 2     & 3     & 3     & 2     & 3     & 3     & \textbf{3} \\
    \midrule
    Total & 14/19 & 44/85 & 57/104 & 40/68 & 65/112 & 74/109 & 64/182 & 41/146 & 68/95 & \textbf{82/101} \\
    \bottomrule
    \end{tabular}%
\end{table*}%

\textbf{Experimental Design.}
In this section, we aim to evaluate the performance of {\toolname}.
We employ the 395 real-world bugs presented in the Defects4J-v1.2 dataset and compare {\toolname} with state-of-the-art APR techniques, including traditional and learning-based ones.
We report the performance of all compared techniques under perfect fault localization (i.e., the ground-truth buggy statement is known to the techniques).

\textbf{Results.}
Table \ref{tab:effectiveness} presents the number of bugs that different APR techniques successfully fix on the Defects4J-v1.2 dataset.
Overall, we find that {\toolname} substantially outperforms the compared APR techniques including both traditional and learning-based APR techniques.
{\toolname} is able to generate correct patches for 82 real-world bugs, 20.59\% (14 bugs), 26.15\% (17 bugs) and 14.8\% (8 bugs) more than TBar, Recoder and AlphaRepair.
In particular, {\toolname} fixes 11, 24, 16, 25, 3, and 3 bugs for Chart, Closure, Lang, Math, Mockito, and Time projects, respectively, four of which are best-performing (bold in Table \ref{tab:effectiveness}).
More importantly, we find that {\toolname} achieves a correct rate of 81.19\% (82/101) for plausible patches, 9.61\% (68/95), 23.15\% (65/112) and 13.30\% (74/109) higher than TBar, Recoder and AlphaRepair, indicating that {\toolname} is able to alleviate the long-standing patch overfitting problem in the community of APR.

\begin{figure}[h]
    \centering
    \includegraphics[width=0.9\linewidth]{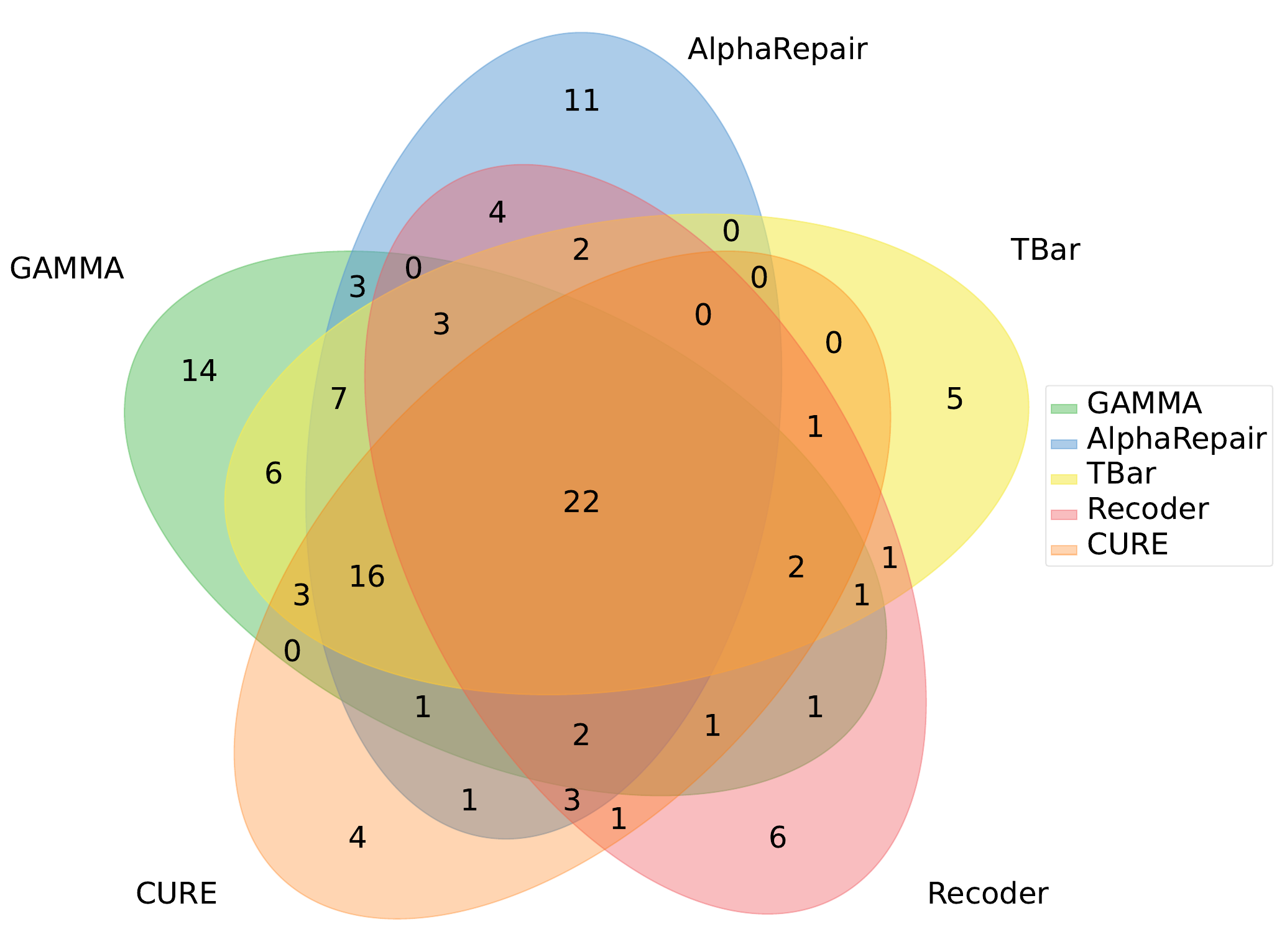}
    \caption{The overlaps of the bugs fixed by different approaches}
    \label{fig:venn}
\end{figure}

\textbf{Overlap Analysis.}
To investigate what extent {\toolname} complements existing APR techniques, we further calculate the number of overlapping bugs fixed by different techniques.
One best-performing traditional technique (i.e., \delete{Tbar}\revise{TBar}) and three best-performing learning-based techniques (i.e., AlphaRepair, CURE, and Recoder) are selected. 
As shown in Fig.~\ref{fig:venn}, {\toolname} fixes 14 unique bugs that other APR approaches fail to fix,  which is 11, 3, 8, and 10 more than TBar, AlphaRepair, CURE, and Recoder, respectively.
More importantly, as a template-based APR technique, there are 22 correctly-fixed bugs unique to {\toolname} compared with \delete{Tbar}\revise{TBar}, highlighting the benefits of mask prediction performed by UniXcoder.
Overall, the results demonstrate that {\toolname} is complementary to these best-performing APR techniques, to increase the number of correctly-fixed bugs in the Defects4J-v1.2 benchmark.

\begin{lstlisting}[float,floatplacement=H,language=diff,caption={The Defects4J bug Math-94},label={listing:math94}]
@@ -409,7 +409,7 @@ public static double factorialLog(final int n) {
public static int gcd(int u, int v) {
-   if (u * v == 0) {            
+   if ((u == 0) || (v == 0)) {
        return (Math.abs(u) + Math.abs(v));
    }
\end{lstlisting}

\textbf{Case Study.}
We have demonstrated the superior performance of {\toolname} over a state-of-the-art template-based tool \delete{Tbar}\revise{TBar}, which is most related to our work.
To further investigate the effectiveness of {\toolname}, we provide some examples of bugs that {\toolname} is able to fix but TBar fails to.
Listing~\ref{listing:math94} presents the bug Math-94 from the Defects4J-v1.2.
Math-94 can be fixed with the fix template of mutating conditional expressions. 
The conditional expression ``u * v == 0'' within an if statement is replaced by ``(u == 0) $||$ (v == 0)''. 
TBar deals with this template in the way of replacing the suspicious expression with other compatible ones collected from the same local file, while {\toolname} directly replaces the expression with a mask token so that they can later be predicted by the mask prediction task from the pre-trained model, making it possible to generate new expressions that can correctly fix the bugs.

\begin{lstlisting}[float,floatplacement=H,language=diff,,caption={The Defects4J bug Closure-52},label={listing:closure52}]
@@ -742,7 +742,7 @@ static boolean isSimpleNumber(String s) {
          return false;
          }
     }
-    return len > 0;
+    return len > 0 && s.charAt(0) != '0';
     }
     static double getSimpleNumber(String s) {
\end{lstlisting}
 
Listing~\ref{listing:closure52} presents a similar example of the bug Closure-52 from the Defects4J-v1.2.
Closure-52 denotes the 52nd buggy version of the Google Closure Compiler project in Defects4J.
To fix this bug, we need to insert a new sub-conditional expression into the original expression.
\delete{Tbar}\revise{TBar} fails to generate a correct patch with  improper operation and variables while {\toolname} is able to directly predict the masked expression with corresponding code context.

\subsection{Generalizability of {\toolname}}

\begin{table*}[htbp]
  \centering
  \caption{Comparison on additional datasets}
    \begin{tabular}{c|ccccccc}
    \toprule
    Project & AlphaRepair & Recoder & CURE  & CoCoNuT & CIRCLE & \delete{Tbar}\revise{TBar}  & {\toolname} \\
    \midrule
    Defects4J 2.0 & 36    & 11    & -     & -     & -     & 8     & 45 \\
    QuixBugs & 28    & 17    & 26    & 13    & 19    & -     & 22 \\
    \bottomrule
    \end{tabular}%
  \label{tab:newbugs}%
\end{table*}%

\textbf{Experimental Design.}
We have demonstrated that {\toolname} achieves impressive performance to repair real-world bugs from the widely-adopted Defects4J-v1.2 benchmark on top of fix patterns.
Durieux et al.~\cite{durieux2019empirical} demonstrate that there exists a common benchmark overfitting phenomenon in APR evaluation, i.e.,  APR tools usually perform significantly better on Defects4J-v1.2 than on other benchmarks.
In this section, according to prior work~\cite{xia2022less,zhu2021syntax}, to evaluate the generalizability of {\toolname}, we continue to conduct some extended experiments on additional projects for further evaluation.

\textbf{Results.}
Table \ref{tab:newbugs} presents the comparison results of {\toolname} against baselines on Defects4J-v2.0 and QuixBugs.
In Defects4J-v2.0, following some recent work \cite{ye2021automated,xia2022less}, we only focus on those bugs whose patches are confined to a single location.
Overall, {\toolname} generates 45 correct patches in the given 257 buggy programs, outperforming both traditional and learning-based approaches.
We find that the performance achieved on the Defects4J-v2.0 dataset is commonly less than that achieved on the Defects4J-v1.2 dataset.
For example, AlphaRepair fixes 18.73\% (74/395) of bugs from  Defects4J-v1.2 while only fixes 14.01\% (36/257) of bugs from Defects4J-v2.0.
Based on our analysis of the two datasets, the possible reason is that Defects4J-v2.0 contains a harder set of projects for APR with a different variety of fixes compared to Defects4J-v1.2.
Despite that, {\toolname} is able to generate 9, 34, and 37 more correct patches, which is the highest number among all approaches.
We also find that as a template-based approach, TBar is able to generate a high amount of correct patches (68) for Defects4J-v1.2, while it only generates a limited number of correct patches (8) for Defects4J-v2.0.
The possible reason may be that most fix patterns are designed to target Defects4J-v1.2, which may not generalize to other unseen projects, such as Defects4J-v2.0;
Besides, learning-based approaches also suffer from moving to a harder evaluation dataset since the code transformation patterns are learned from training datasets which might not be present in Defects4J-v2.0. 
In contrast, {\toolname} is able to address the generalizability issue without training on specific bug datasets, which makes it less prone to suffer from generalizability issues of traditional template-based or learning-based tools.

Apart from Defects4J-v2.0, we also try to validate our approach on QuixBugs, which extracts bugs from Quixey Challenge and translates them into both Java and Python languages. 
Since our fix templates are designed for Java, we only focus on Java programs in QuixBugs following previous work~\cite{liu2019tbar}.
Table \ref{tab:newbugs} shows that among 40 bugs in QuixBugs, 22 are correctly fixed by {\toolname}, highlighting the competitive performance of {\toolname} against state-of-the-art approaches.
It is worth noting that most of the templates are summarized from Defects4J-v1.2, which may mean that some templates cannot be applied to any bugs except those from Defects4J-v1.2. 
Thus, {\toolname} may be limited by the lack of more efficient fix templates when coming to other new bug datasets.
For example, although various types of templates along with sub-templates are defined, some of the templates cannot be used to fix at least one bug from QuixBugs. 
As a result, we expect to explore more general fix templates in the future to further improve the performance of template-based APR.

\subsection{Scalability of {\toolname}}
\label{sec:rq3}

\textbf{Experimental Design.}
To further investigate whether the performance of {\toolname} is affected by different pre-trained models, we apply two other advanced models to perform the mask prediction task: CodeBERT and ChatGPT.
CodeBERT \cite{feng2020codebert} is a pre-trained model for programming and natural languages, and mask prediction is one of its pre-training tasks. 
ChatGPT is a state-of-the-art language model that has shown impressive ability in conversations with human beings.
We also use Defects4J-v1.2 as a benchmark but replace UniXcoder with these two models in the process of filling masks to find out to what extent pre-trained models influence the effectiveness of template-based program repair.

Similar to UniXcoder, CodeBERT can also generate predictions for a mask token ``$<$mask$>$'' in the given code snippet. 
The difference between them is that UniXcoder can predict several continuous tokens for a single mask while CodeBERT can only give one token for a mask. 
However, there is usually more than one token under a mask, so when using CodeBERT, we have to use different numbers of successive masks to mask the initial code and then predict them sequentially.
\revise{We do not know the exact number of masks we should use (i.e., the number of tokens in the fixed code) as there could be a great many possibilities in the patch. So while masking, we naively try all the mask numbers from 1 to 20, which is a range suitable to most cases.}
In every iteration, a mask is predicted and a joint score for each prediction is calculated. 
Those predictions with the highest scores will be chosen to replace the mask, and the next mask will be predicted according to the previous predictions. 
We set the beam size as 250, the same as that used in UniXcoder, so in each iteration, CodeBERT will give 250 most possible predictions for the mask.
Taking the process of fixing the Defects4J Closure-10 bug as an example (shown in Listing~\ref{listing:closure10}), the patch of the bug involves a change in the method name.
To fix this bug, the method name \emph{allResultsMatch} is replaced with masks, and then CodeBERT is asked to give 250 predictions for the first mask. 
Among the predictions, the token ``any'' has a relatively high score, so it replaces the first mask and CodeBERT will continue to predict the next mask until all masks are filled.

Different from UniXcoder, ChatGPT is fine-tuned from GPT-3.5 and close-sourced.
We can access ChatGPT with ChatGPT’s API of gpt-3.5-turbo-0301, which is the latest version available.
We interact with ChatGPT through natural language conversations, i.e., sending requests to ChatGPT or receiving responses from ChatGPT.
To fill the masks with ChatGPT, we first give it some prompts, instructing it to return back predictions for the mask. 
Following the prompts we then add the masked buggy line along with its context to form the complete query for ChatGPT. 
In our experiment, the input for ChatGPT starts with a prompt ``\textit{Next token prediction task, the first line is a comment to help prediction, just return 250 possible predictions for $<$mask$>$ with highest probability: }'', and then the bug context we give is the same as the input for UniXcoder, which consists of a commented buggy line and the whole method where the buggy line belongs.
\revise{Besides, due to the benefits of the designed prompt, we do not set the number of masked tokens in the buggy code, which is the same as UniXcoder.}

\begin{figure}[htbp]
    \centering    \includegraphics[width=0.9\linewidth]{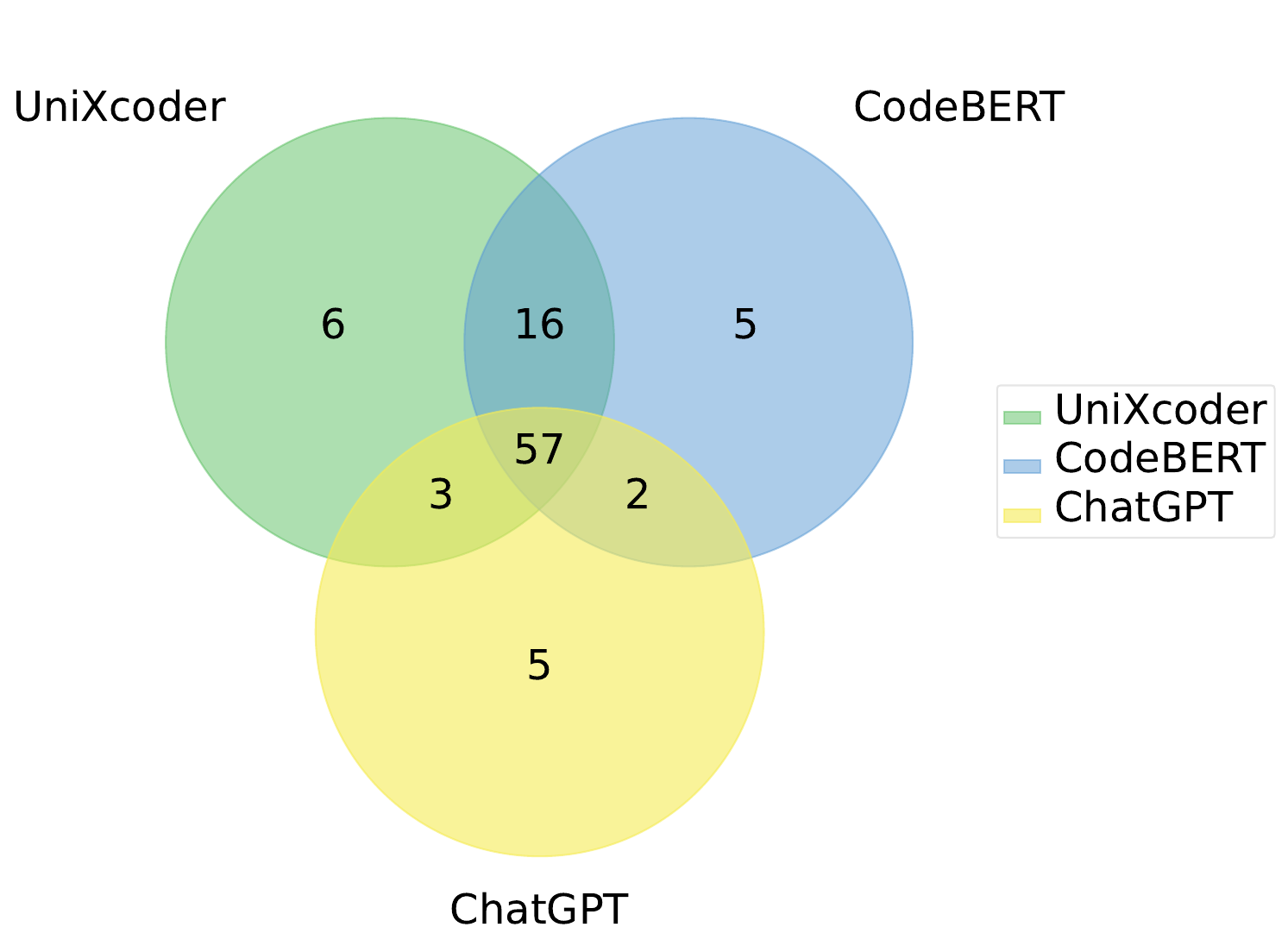}
    \caption{The overlaps of the bugs fixed by different pre-trained language models}
    \label{fig:model}
\end{figure}

\textbf{Results.}
Fig.~\ref{fig:model} presents the repair results of different pre-trained language models.
Overall, the combination of the three models is able to fix 93 bugs from Defects4J-v1.2, demonstrating these models can be used together by {\toolname} to further increase the number of correct patches that can be generated.
In particular, we find when using CodeBERT to perform the mask prediction task, 80 bugs in total are fixed by {\toolname} correctly, only two less than the bugs that {\toolname} with UniXcoder fixes. 
However, it takes much more time for CodeBERT to generate correct patches, as the number of masks that should be used in fixing a bug is unpredictable, and we have to run the mask prediction program on the same bug and the same fix template for a lot of times, each with a different mask number. 
In contrast, UniXcoder circularly predicts the next token for a mask until an EOF token is generated, so only one mask is required to fix the bug.
We also find that {\toolname} with ChatGPT only fixes 67 bugs correctly, performing not well as {\toolname} with UniXcoder and CodeBERT in mask prediction.
The possible reason lies in that UniXcoder is pre-trained with a masked language modeling objective, where some training text is artificially masked and the training objective is to predict the real text.
However, ChatGPT is designed for natural language conversations and it is unclear how ChatGPT is pre-trained due to it being close-sourced.
Thus, it is natural to employ UniXcoder to recover masked code tokens for buggy code snippets in our approach.
Future researchers should further explore how to betters utilize ChatGPT (e.g., designing other prompts) for mask prediction.

\section{Threats to Validity}
\label{sec:threats}

The first threat to validity comes from the manual inspection of correct patches.
To alleviate the influence of potential bias,  following previous works \cite{yuan2022circle,jiang2021cure}, three authors manually verify all plausible patches (i.e. patches that successfully pass the test) based on ground truth patches (i.e., developer patches).
A plausible patch is considered to be correct if all three authors identify it as equivalent to a ground truth patch semantically.
To facilitate the replication and verification of our experiments, we also release the relevant materials (including all source code and correct patches) publicly~\cite{myurl}.

\revise{
The second threat to validity is the fault localization setting.
We evaluate {\toolname} and baselines on perfect fault localization (i.e., ground-truth buggy location is known) due to two reasons.
First, off-the-shelf fault localization approaches affect the performance of APR techniques significantly, introducing a bias in comparison results~\cite{liu2019you}.
The perfect fault localization mitigates the influence of differences in different fault localization approaches on the repair results and enables a fair assessment of repair performance independently of the fault localization approach used.
Second, the most recent APR techniques ~\cite{yuan2022circle,jiang2023impact,jiang2021cure,lutellier2020coconut,xia2022practical} are only evaluated with perfect fault localization, which makes our work also use perfect localization to ensure direct comparison.
However, this comparison setting may bring bias in repair performance since the perfect fault localization results are usually unavailable in real practice.
Despite that, we believe that perfect fault localization has little impact on our results, as perfect localization can show the pure performance of different APR approaches.
In the future, we attempt to report the repair performance of {\toolname} and baselines with both the automated fault localization (e.g., Ochiai~\cite{wong2016survey}) and perfect fault localization.

}

\revise{The third threat to validity comes from the potential of data leakage of pre-trained models.
In our experiment, we implement {\toolname} on top of UniXCoder and evaluate {\toolname} on the widely-adopted benchmark Defects4J-v1.2.
Considering that UniXCoder is pre-trained with millions of code snippets, there may exist some bugs in the evaluation benchmark of Defects4J-v1.2 that appear in the pre-training dataset of UniXcoder.
We perform a manual inspection to check whether the fixed bugs by {\toolname} are leaked into  the pre-training dataset.
In particular, we query the pre-training datasets including 2.3M functions paired with comments and 4.1M unimodal code from CodeSearchNet.
The manual inspection is performed by two authors independently and confirmed by a third author. 
We find there are three bugs leaked into the pre-training set, i.e., Closure-73, Closure-126, and Time-19.
For the three bugs, we manually perturb the buggy code (e.g., changing variable names, adding dead code) and find GAMMA is still able to generate the correct patches for all three bugs. We also find that if we exclude the three overlapping bugs, GAMMA still outperforms state-of-the-art APR techniques (79 vs. 68 for TBar, 79 vs. 74 for AlphaRepair). 
Thus, we are confident that the data leakage is not a key point to our conclusion.
}

\section{Related Work}
\subsection{Automated Program Repair}
Existing APR techniques can be divided into four categories, 
i.e., heuristic-based~\cite{ Martinez2016Astor,Yuan2018Arja}, constraint-based~\cite{Durieux2016Dynamoth,Xuan2016Nopol, mechtaev2016angelix}, template-based \cite{koyuncu2020fixminer,Liu2019Avatar,liu2019tbar} and learning-based repair techniques~\cite{li2020dlfix,zhu2021syntax,lutellier2020coconut}.
Our work is related to template-based and learning-based APR, discussed as follows.

Template-based APR, which generates patches with the help of fix patterns, represents state-of-the-art among traditional APR techniques.
TBar~\cite{liu2019tbar} systematically collects and summarizes fix patterns from previous literature and investigates the effectiveness of patch generation based on these templates. 
Some other techniques explore fix patterns in various ways. 
For example, 
PAR~\cite{kim2013automatic} manually extracts fix patterns from 60,000 human-written patches.
FixMiner~\cite{koyuncu2020fixminer} mines fix patterns with an iterative clustering strategy. 
AVATAR~\cite{Liu2019Avatar} leverages fix patterns from static bug detection tools to generate patches. 
Different from most existing template-based APR techniques that focus on mining fix patterns, {\toolname} is the first work that aims to address the donor code issue by integrating pre-trained models with a fill-in-the-blank task.

With large available open-source code corpora, learning-based APR, which applies machine learning to the bug-fixing objective, is getting growing attention.
For example, DLFix~\cite{li2020dlfix} uses a tree-based recurrent neural network (RNN) model to learn from bug fixes and surrounding contexts in the form of an abstract syntax tree. CoCoNuT~\cite{lutellier2020coconut} introduces a novel context-aware neural machine translation (NMT) architecture that separately represents the buggy code and context.
CURE~\cite{jiang2021cure} attempts to break the limit of existing NMT-based techniques by pre-training a programming language model on a large codebase, introducing a new code-aware search strategy, and using subword tokenization to narrow the search space.
Recoder~\cite{zhu2021syntax} is a syntax-guided edit decoder with placeholder generation, which provides a novel provider/decider architecture to guarantee that patches with correct syntax are generated.
Different from existing learning-based APR techniques generating patches from scratch with bug-fixing data training, {\toolname} aims to directly predict correct code tokens with the help of fix patterns in a zero-shot scenario.

\revise{Recently, there exists an increasing number of APR techniques on top of pre-trained models.
For example, Yuan et al.~\cite{yuan2022circle} propose CIRCLE, a T5-based program repair framework equipped with continual learning ability across multiple languages.
Xia et al.~\cite{xia2022less} propose AlphaRepair, a cloze-style APR approach based on CodeBERT without fine-tuning on historical bug-fixing data.
In our work, we include CIRCLE and AlphaRepair as baselines in the experiment.
Sobania et al.~\cite{sobania2023analysis} investigate the performance of ChatGPT on the QuixBugs benchmark.
Mashhadi et al. \cite{mashhadi2021applying} investigate the performance of fine-tuning CodeBERT to fix software bugs from ManySStuBs4J.
Jiang et al.~\cite{jiang2023impact} explore the performance of pre-trained models with and without fine-tuning for the program repair domain.
Xia et al.~\cite{xia2022practical} further present an extensive evaluation of recent pre-trained models for fixing real-world projects and find state-of-the-art pre-trained models are able to fix a considerable number of bugs.
For example, CodeX, the most effective one, fixes 99 bugs in Defects4J-v1.2 with a total combination of three repair settings.
We exclude CodeX as a baseline in our experiment due to the uncertainty of training data in such black-box large pre-trained models.
}

\subsection{Pre-trained Language Models and Applications}
\label{sec:premodel}

In this section, we introduce some typical pre-trained language models and then discuss the applications of pre-trained language models to some code-related tasks, e.g., code search.

\subsubsection{Pre-trained Language Models}

Pre-trained language models have shown promising results on NLP tasks. 
BERT \cite{kenton2019bert} is a model to condition on left and right contexts in all layers so as to pre-train deep bidirectional representations from unlabeled text.
GPT-3~\cite{brown2020gpt} is an autoregressive language model having 175 billion parameters, significantly outnumbering the parameters in previous language models. 
ChatGPT~\cite{chatgpt} is the currently most popular language model fine-tuned from GPT-3
and is receiving attention from both scientific and industrial fields. 
The most remarkable feature of ChatGPT is that it can generate human-like responses and communicate with human beings like what a real human can do.

Inspired by the success of pre-trained models in NLP, many researchers apply the pre-trained model to code-related tasks.
Feng et al. \cite{feng2020codebert} propose a bimodal pre-trained model (CodeBERT) for both programming language and natural language. CodeBERT is developed on Transformer-based neural architecture and pre-trained with the task of masked language modeling, which is to predict tokens, and replaced token detection.
Guo et al. \cite{guo2022unixcoder} present UniXcoder, a unified cross-modal pre-trained model for programming language.
UniXcoder utilizes mask attention matrices with prefix adapters to control the behavior of the model and leverages cross-modal contents such as AST and code comment to enhance code representation. 
Different from these studies designing novel pre-train models from scratch, we attempt to boost template-based APR on top of these pre-trained models.

\subsubsection{Applications of Pre-trained Models}
\label{sec:other_se}

In addition to the above-mentioned typical pre-trained models, researchers have also applied such pre-trained models to some code-related domains (e.g., code completion, and program repair). 
\revise{Mastropaolo~et al.~\cite{mastropaolo2022using} present an empirical study to investigate the usage of pre-trained models for four code-related tasks, including program repair, mutants injection, assertion generation, and code summarization.
A similar strategy combining mutation patterns and pre-trained models is adopted in mutation testing.
For example, Degiovanni et al.~\cite{degiovanni2022mubert} introduce $\mu$BERT, a CodeBERT-based mutation testing tool by masking a token from the expression and replacing the masked token with the predicted one from CodeBERT.
Richter~et al.~\cite{richter2022learning} propose a contextual mutation operator by employing CodeBERT to produce a context-dependent distribution over feasible token replacements.}
Recently, Zhang et al.~\cite{zhang2023pre} conduct an extensive empirical study to investigate the performance of pre-trained models in repairing security vulnerabilities and propose a enhanced approach with bug fixing transfer learning.
Although there exist some SE tasks (e.g., mutation testing and program repair) benefitting from pre-trained models, in this work, we perform the first work to employ pre-trained models to directly predict the correct code with the help of fix patterns.

\section{Conclusion}
In this work, we present {\toolname}, an innovative template-based APR tool that assimilates the advances of fix templates and pre-trained models. 
{\toolname} first defines a set of mask fix templates by masking buggy code tokens with corresponding code context.
{\toolname} then uses the off-the-shelf pre-trained models to directly recover the correct code with a mask prediction task.
More importantly, {\toolname} can be built on various pre-trained models under a zero-shot learning setting and we implement it as a practical APR tool using the recent UniXcoder model.
The experimental results on the popular Defects4J-v1.2 dataset have shown promising performance, e.g., 82 bugs are fixed by {\toolname}, outperforming all state-of-the-art APR techniques.
We also demonstrate that {\toolname} is able to address the dataset overfitting well, e.g., 45 and 22 bugs are fixed in Defects4J-v2.0 and Quixbugs.
We further demonstrate that {\toolname} is generalizable to different pre-trained language models, such as CodeBERT and ChatGPT.

\section*{Acknowledgment}
The authors would like to thank the anonymous reviewers for their insightful comments. This work is supported partially by the National Natural Science Foundation of China (61932012, 62141215).

\ifCLASSOPTIONcaptionsoff
  \newpage
\fi

\bibliographystyle{IEEEtran}
\bibliography{reference}

\end{document}